\documentclass[aps,prb,twocolumn,superscriptaddress,amsmath,amssymb,floatfix]{revtex4-2}

\usepackage{graphicx}
\usepackage{dcolumn}
\usepackage{bm}
\usepackage{hyperref}
\usepackage{booktabs}
\usepackage{etoolbox}

\hypersetup{
    colorlinks=true,
    linkcolor=blue,
    citecolor=blue,
    urlcolor=blue
}

\begin{document}

\title{Resolving Spectral Gaps of the Frustrated Spin-1/2 $J_1$-$J_2$ Chain via Dual Spin-Sector Variational Quantum Eigensolver}

\author{Sk Saniur Rahaman}
\email{sksrahaman@imsc.res.in}
\email{saniur1994@gmail.com}
\affiliation{The Institute of Mathematical Sciences, A CI of Homi Bhabha National Institute, Chennai 600113, India}

\date{\today}

\begin{abstract}
Finding excited states and resolving spectral gaps in frustrated quantum spin systems using Variational Quantum Eigensolvers is complicated by state-ordering inversions. In the frustrated spin-1/2 $J_1$-$J_2$ Heisenberg chain, standard single-sector variational quantum deflation collapses onto intermediate triplet excitation, preventing the isolation of excited singlets and the singlet-singlet gap. To resolve these gaps, a Dual Spin-Sector Variational Quantum Eigensolver framework is introduced for near-term quantum architectures. By partitioning the search across magnetization sectors $S^z=0$ and $S^z=1$, the lowest triplet is resolved as an unconstrained ground state in $S^z=1$ with zero deflation penalty. Leveraging exact global spin-rotation symmetry, this state is mapped analytically into $S^z=0$ without ancilla overhead, unlocking dual deflation via shallow compute-uncompute circuits. Tracking both singlet-triplet and singlet-singlet gaps across coupling ratios $J_2/J_1$ from $0$ to $1$ for chains up to $16$ qubits faithfully captures all critical phenomena: the continuous Berezinskii-Kosterlitz-Thouless transition, exact Majumdar-Ghosh dimer degeneracy, level-ordering inversion in the gapless liquid phase, and incommensurate spiral excitations. Circuit depth scaling up to twenty layers provides an overparameterization mechanism overcoming classical local minima in strongly frustrated regimes, reducing relative energy errors below $1\%$ and elevating state fidelities above $97\%$ on $16$ qubits. Projective measurement sampling demonstrates resilience against shot noise, providing a practical blueprint for exploring frustrated quantum magnetism on near-term quantum devices.
\end{abstract}

\maketitle

\section{Introduction}
\label{sec:intro}
Frustrated quantum spin systems have garnered immense interest in condensed matter physics and quantum information science owing to their rich quantum phases and exotic phenomena driven by competing antiferromagnetic exchange interactions~\cite{diep2004frustrated, Lhuillier2002}. However, resolving the ground and low-lying excited states of these systems remains a formidable challenge. Conventional classical computational methods face severe intrinsic limitations when applied to frustrated quantum magnets. Due to the exponential growth of the Hilbert space, exact diagonalization (ED) is restricted to small system sizes. Quantum Monte Carlo (QMC) simulations are severely impeded by the notorious sign problem induced by geometric frustration. While the Density Matrix Renormalization Group (DMRG)~\cite{White1992} and Matrix Product States (MPS) represent premier methods for quasi-one-dimensional geometries, accurately capturing excited-state manifolds and dense spectral gaps remains computationally demanding. Neural Quantum States (NQS) have recently emerged as expressive variational wave functions, yet their training in strongly frustrated regimes remains computationally expensive~\cite{Carleo2017}. 

To circumvent the limitations of classical computing, hybrid classical-quantum algorithms tailored for Noisy Intermediate-Scale Quantum (NISQ) processors have emerged as a prominent pathway for quantum simulation~\cite{Preskill2018}. In this context, the Variational Quantum Eigensolver (VQE) has established itself as a leading variational framework~\cite{Peruzzo2014, McClean2016, Tilly2022}. A variety of parameterized quantum circuit ans\"atze have been developed to simulate magnetic Hamiltonians on quantum hardware. Uvarov et~al.~\cite{PRB102_075104} evaluated VQE architectures on frustrated spin-fermion chains and analyzed fermion-to-qubit mappings. Feulner and Hartmann~\cite{PRB106_144426} designed hardware-efficient circuits that eliminate diagonal SWAP gates, while Sheils and Rhone~\cite{Sheils2024} investigated frustrated quantum spin phase transitions. Kattem\"{o}lle and van Wezel~\cite{PRB106_214429} introduced symmetry-preserving VQE protocols for frustrated kagome geometries, and Kirmani et~al.~\cite{Kirmani2025} demonstrated hardware implementations of frustrated Ising models, building upon earlier pioneering demonstrations on superconducting and trapped-ion processors~\cite{Wecker2015, Kandala2017}. Adaptive variational strategies such as ADAPT-VQE~\cite{Grimsley2019} dynamically optimize circuit depth, while theoretical investigations by Cerezo et~al.~\cite{Cerezo2021} proved that cost-function locality is vital for evading Barren plateaus~\cite{McClean2018} in shallow circuits. Nevertheless, existing VQE studies of frustrated quantum spin chains focus overwhelmingly on ground-state energies ($E_S^0$), leaving state-ordering inversions in excitation spectra largely unaddressed.

Although full Quantum Phase Estimation (QPE)~\cite{AspuruGuzik2005} requires deep, fault-tolerant circuits beyond current experimental capabilities, several excited-state variational protocols have emerged, including Variational Quantum Deflation (VQD)~\cite{Higgott2019}, Subspace-Search VQE (SSVQE)~\cite{Nakanishi2019}, real-time evolution subspace methods~\cite{Jones2019}, the quantum equation of motion (qEOM)~\cite{Ollitrault2020}, Quantum Filter Diagonalization~\cite{Parrish2019}, and Quantum Krylov algorithms~\cite{Cortes2022}. In particular, VQD enforces state orthogonality by introducing overlap penalty terms into the cost function to project out previously determined lower-lying states. However, in frustrated quantum spin chains, standard single-sector deflation encounters a fundamental bottleneck: the physical level ordering of spin sectors causes variational optimization within $\mathcal{H}_{S^z=0}$ to collapse onto intermediate triplet state ($|\psi_T^0\rangle$) rather than resolving true excited singlet ($|\psi_S^1\rangle$), impeding accurate extraction of low-lying spectral gaps.

The frustrated spin-1/2 $J_1$-$J_2$ Heisenberg chain, featuring nearest-neighbor coupling $J_1$ and next-nearest-neighbor coupling $J_2$, serves as a prototypical benchmark system exhibiting rich quantum phases and transitions governed by competing exchange interactions. Foundational numerical, field-theoretic, and DMRG investigations of this model have comprehensively characterized its ground-state dimer order, level crossings, bond-order wave physics, and incommensurate spiral correlations~\cite{White1996, White2007, Chitra1995, Pati1996, Kumar2010bow, Kumar2010zigzag, Kumar2012parity, Parvej2015, Parvej2017, Kumar2016yjunc, Soos2016, JPCM_35_145801, Haldar2024bkt}. However, accurately calculating excited-state spectra, low-lying spectral gaps, and state-ordering crossovers remains computationally challenging on classical architectures due to rapid entanglement growth and high-dimensional spectral overlaps~\cite{Chandra1988, Dagotto1989}. To address this challenge, this work develops a Dual Spin-Sector Parameterized VQE framework formulated for NISQ processors. By partitioning the variational search across distinct magnetization sectors ($S^z=0$ and $S^z=1$), mapping states via exact global $SU(2)$ spin-lowering ($[H, S^-]=0$), and evaluating dual overlap penalties through shallow Compute-Uncompute circuits, the framework resolves both singlet and triplet low-lying excitations across the complete phase diagram without deep symmetry-projection circuits or ancilla overheads. All variational algorithms reported herein are evaluated via high-precision classical state-vector simulations and exact parameter gradients, complemented by realistic projective measurement shot-noise sampling ($N_{\text{shots}} \in [10^3, 10^6]$) to simulate NISQ execution conditions while cleanly isolating variational ansatz expressibility from physical processor calibration drift and decoherence. Numerical benchmarks from $L=8$ up to $L=16$, combined with circuit depth scaling beyond $p=L$, demonstrate high accuracy and numerical stability across the gapless, dimerized, and incommensurate regimes.

The remainder of this paper is organized as follows. Section~\ref{sec:model} introduces the frustrated $J_1$-$J_2$ Heisenberg chain and its low-energy phase diagram. Section~\ref{sec:method} details the Dual Spin-Sector VQE formulation, ansatz architecture, and algorithmic workflow. Section~\ref{sec:results} presents variational benchmarks across phase boundaries, scalability analyses for $L=16$, circuit depth scaling beyond $p=L$, and NISQ shot-noise resilience. Section~\ref{sec:discussion} synthesizes the findings and discusses the hardware co-design outlook.

\section{Model Hamiltonian and Quantum Phases}
\label{sec:model}
\subsection{Model Hamiltonian}
The Hamiltonian of the one-dimensional frustrated spin-1/2 $J_1$-$J_2$ Heisenberg chain with periodic boundary conditions (PBC) is given by:
\begin{equation}
    H = J_1 \sum_{i=1}^L \vec{S}_i \cdot \vec{S}_{i+1} + J_2 \sum_{i=1}^L \vec{S}_i \cdot \vec{S}_{i+2},
    \label{eq:hamiltonian_1d}
\end{equation}
where $\vec{S}_i = \frac{1}{2} (\sigma_i^x, \sigma_i^y, \sigma_i^z)$ denotes the spin-1/2 vector operator acting on qubit $i$ and $\sigma_i^{\alpha}$ are the pauli spin operators. $J_1, J_2 > 0$ represent the nearest-neighbor (NN) and next-nearest-neighbor (NNN) antiferromagnetic exchange couplings, respectively. The spin scalar product expands as:
\begin{equation}
    \vec{S}_i \cdot \vec{S}_j = \frac{1}{4} \left( \sigma_i^x \sigma_j^x + \sigma_i^y \sigma_j^y + \sigma_i^z \sigma_j^z \right).
\end{equation}
Because the exchange interactions are isotropic, the Hamiltonian commutes with both the total magnetization operator along the quantization axis, $S^z = \sum_{i=1}^L S_i^z$ ($[H, S^z] = 0$), and the total spin-squared Casimir operator, $S^2 = (\sum_{i=1}^L \vec{S}_i)^2$ ($[H, S^2] = 0$).

\subsection{Quantum Phases and Phase Boundaries}
The low-energy spectrum of Eq.~(\ref{eq:hamiltonian_1d}) is governed by three primary eigenstates:
(i)~The Ground State Singlet $|\psi_S^0\rangle$, possessing total spin $S=0$ and magnetization $S^z=0$, with energy $E_S^0$.
(ii)~The lowest Triplet excitation, possessing total spin $S=1$, which forms a three-fold degenerate multiplet across magnetic quantum numbers $S^z \in \{0, +1, -1\}$, denoted respectively by $|\psi_T^0\rangle$, $|\psi_T^1\rangle$, and $|\psi_T^2\rangle$ with common energy $E_T$. The $S^z=0$ component is $|\psi_T^0\rangle$, while the $S^z=+1$ and $S^z=-1$ components represent fully polarized configurations ($|\psi_T^1\rangle$ and $|\psi_T^2\rangle$).
(iii)~The first Excited Singlet $|\psi_S^1\rangle$, residing strictly within the $S=0, S^z=0$ sector, with energy $E_S^1$.

As the coupling ratio $J_2/J_1$ varies from $0.00$ to $1.00$, the 1D $J_1$-$J_2$ chain traverses three fundamental physical regimes:

\textit{(1)~Gapless Tomonaga-Luttinger Liquid (TLL) Phase ($0 \le J_2/J_1 < 0.2411$)}: The system is gapless in the thermodynamic limit ($\Delta_{ST} = E_T - E_S^0 \to 0$ as $L \to \infty$). Within the zero-magnetization sector $S^z=0$, the triplet state $|\psi_T^0\rangle$ lies energetically below the first excited singlet $|\psi_S^1\rangle$, establishing the state-ordering inversion $E_S^0 < E_T < E_S^1$.
\begin{figure*}[t]
    \centering
    \includegraphics[width=0.90\linewidth,height=0.40\linewidth]{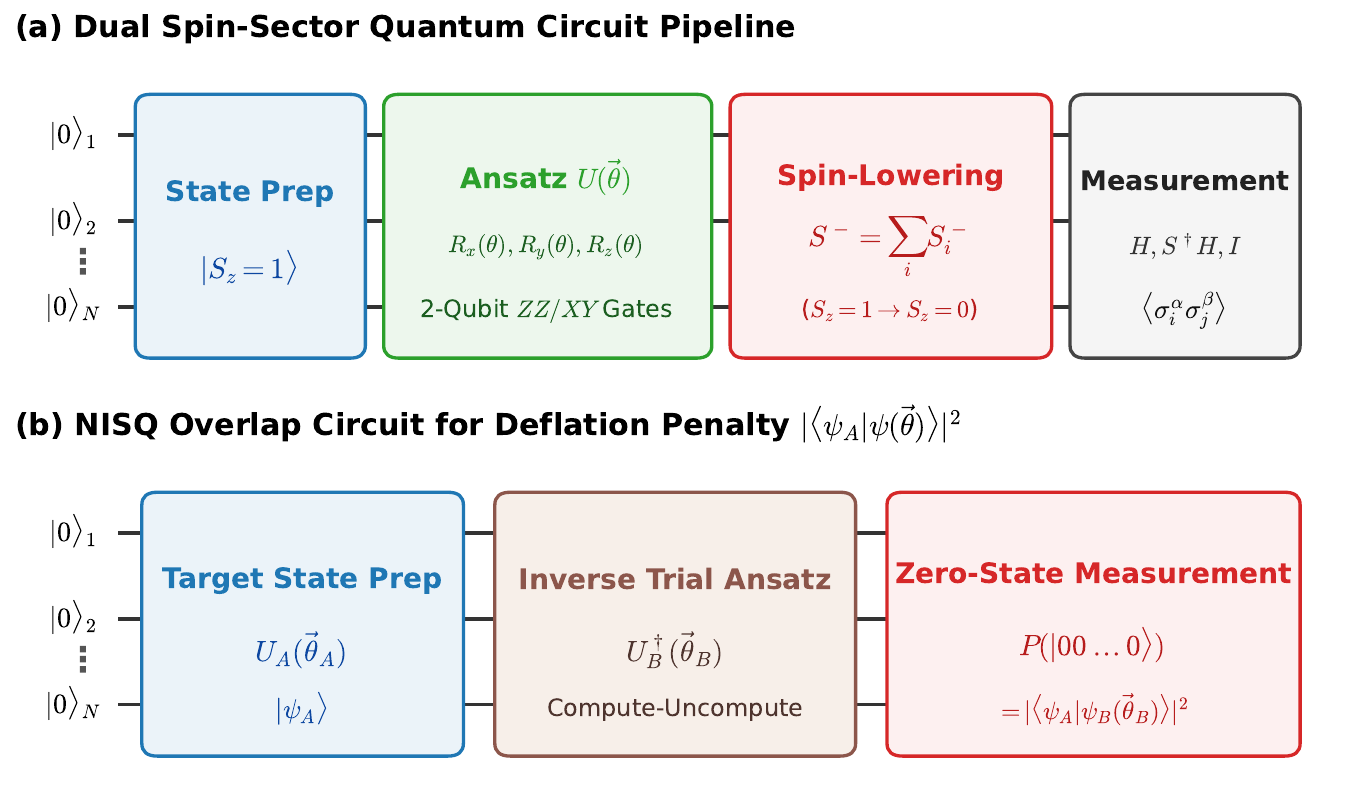}
    \caption{Dual Spin-Sector VQE circuit architecture. (a)~Parameterized ansatz $U(\vec{\theta})$ for magnetization-sector state preparation ($S^z=1$ and $S^z=0$), exact $S^-$ spin-lowering mapping, and Pauli expectation value measurement. (b)~Compute-Uncompute circuit $U_A^\dagger U(\vec{\gamma})$ for evaluating state overlap penalties without ancilla overhead.}
    \label{fig:circuit_pipeline}
\end{figure*}

\textit{(2)~Gapped Dimer Phase and Majumdar-Ghosh Point ($0.2411 \le J_2/J_1 \le 0.50$)}: At the continuous BKT critical point $(J_2/J_1)_c \approx 0.2411$~\cite{Okamoto1992, Chitra1995}, spontaneous dimerization occurs, opening a finite spin gap $\Delta_{ST} > 0$. At the Majumdar-Ghosh point $J_2/J_1 = 0.50$~\cite{Majumdar1969}, the ground state becomes strictly two-fold degenerate, spanned by exact valence-bond product dimer configurations $|\psi_{\text{MG}}^{(1)}\rangle = [1,2][3,4]\dots[L-1,L]$ and $|\psi_{\text{MG}}^{(2)}\rangle = [2,3][4,5]\dots[L,1]$, where $[i,j] = \frac{1}{\sqrt{2}}(|\!\uparrow_i\downarrow_j\rangle - |\!\downarrow_i\uparrow_j\rangle)$ denotes a singlet valence bond between sites $i$ and $j$. At this point, the ground-state energy per site is known analytically to be $E_S^0/L = -3/8 = -0.3750$, and the Singlet-Singlet gap vanishes identically ($\Delta_{SS} = E_S^1 - E_S^0 = 0.0000$), while the spin gap remains robustly open ($\Delta_{ST} = E_T - E_S^0 > 0$).

\textit{(3)~Incommensurate Spiral Phase ($J_2/J_1 > 0.50$)}: Short-range spin-spin correlation functions exhibit oscillatory spatial modulation with an incommensurate pitch vector $q = \arccos(-J_1 / 4J_2)$~\cite{Eggert1996, JPCM_35_145801, Soos2016}. On finite periodic chains, quantum tunneling between dimer patterns forms a quasi-degenerate ground doublet $\{|0^+\rangle, |0^-\rangle\}$ with momentum $k=0$ and $k=\pi$, split by an exponentially small finite-size tunneling gap $\delta E \sim 10^{-3}$. As $J_2/J_1 \to 1.00$, the chain decouples into two weakly-interlaced sublattices of size $L/2$.

\section{Methodology}
\label{sec:method}
\subsection{Dual Spin-Sector VQE Formulation}
To circumvent the state-ordering inversion in $S^z=0$ ($E_S^0 < E_T < E_S^1$), the variational search is partitioned across distinct magnetization sectors $\mathcal{H}_{S^z=0}$ and $\mathcal{H}_{S^z=1}$ [Fig.~\ref{fig:circuit_pipeline}(a)]. The ground singlet state $|\psi_S^0\rangle$ is resolved by minimizing a parameterized quantum circuit $U(\vec{\theta}_S^0)$ in the zero-magnetization sector:
\begin{equation}
    E_S^0 = \min_{\vec{\theta}_S^0} \langle 0 | U^\dagger(\vec{\theta}_S^0) H U(\vec{\theta}_S^0) | 0 \rangle_{S^z=0} \implies |\psi_S^0\rangle.
\end{equation}
Because singlet states ($S=0$) cannot exist within $\mathcal{H}_{S^z=1}$, minimizing a separate ansatz $U(\vec{\theta}_T^1)$ in $S^z=1$ directly isolates the lowest triplet state without singlet contamination:
\begin{equation}
    E_T = \min_{\vec{\theta}_T^1} \langle \psi_T^1(\vec{\theta}_T^1) | H | \psi_T^1(\vec{\theta}_T^1) \rangle_{S^z=1} \implies |\psi_T^1\rangle.
\end{equation}

For the isotropic Heisenberg Hamiltonian [Eq.~(\ref{eq:hamiltonian_1d})], total spin-lowering $S^- = \sum_{i=1}^L \frac{1}{2}(\sigma_i^x - i \sigma_i^y)$ satisfies $[H, S^-] = 0$ exactly due to global $SU(2)$ symmetry. Applying $S^-$ maps $|\psi_T^1\rangle$ into $S^z=0$:
\begin{equation}
    |\psi_T^0\rangle = \frac{S^- |\psi_T^1\rangle}{\| S^- |\psi_T^1\rangle \|},
    \label{eq:s_minus}
\end{equation}
yielding the exact degenerate triplet eigenstate in $S^z=0$ with energy $E_T$ without extra variational parameters. On quantum hardware, the non-unitary operator $S^-$ is evaluated using single-qubit Pauli observables via the total spin-raising operator $S^+ = \sum_{i=1}^L \frac{1}{2}(\sigma_i^x + i \sigma_i^y)$:
\begin{align}
    &|\langle \psi_T^0 | \psi(\vec{\gamma}) \rangle|^2 = \frac{|\langle \psi_T^1 | S^+ | \psi(\vec{\gamma}) \rangle|^2}{\langle \psi_T^1 | S^+ S^- | \psi_T^1 \rangle} \nonumber \\
    &\quad= \frac{1}{8} \left| \sum_{i=1}^L \langle 0 | U_T^\dagger(\vec{\theta}_T^1) (\sigma_i^x + i\sigma_i^y) U(\vec{\gamma}) | 0 \rangle \right|^2,
    \label{eq:s_minus_hardware_derivation}
\end{align}
where the denominator evaluates analytically to $\langle 1, 1 | S^+ S^- | 1, 1 \rangle = 2$. Transition amplitudes are measured using shallow Compute-Uncompute circuits [Fig.~\ref{fig:circuit_pipeline}(b)].

To resolve the true excited singlet $|\psi_S^1\rangle$ within $S^z=0$, a dual-sector penalty cost function is minimized:
\begin{align}
    \mathcal{L}(\vec{\gamma}) &= \langle \psi(\vec{\gamma}) | H | \psi(\vec{\gamma}) \rangle + \beta_1 |\langle \psi_S^0 | \psi(\vec{\gamma}) \rangle|^2 \nonumber \\
    &\quad + \beta_2 |\langle \psi_T^0 | \psi(\vec{\gamma}) \rangle|^2,
    \label{eq:dual_cost}
\end{align}
where $|\psi(\vec{\gamma})\rangle = U(\vec{\gamma})|0\rangle_{S^z=0}$, and $\beta_1 = \beta_2 = 2.0 \cdot |E_S^0|$ are penalty weights set above the low-energy gap. Because $[H, S^2] = 0$, $|\psi_S^0\rangle$ ($S=0$) and $|\psi_T^0\rangle$ ($S=1$) are mutually orthogonal by $SU(2)$ symmetry, guaranteeing robust convergence to the physical excited singlet $|\psi_S^1\rangle$ with energy $E_S^1$.

\subsection{Hamiltonian Variational Ans\"atze and Quantum Circuit}
For NISQ execution, parameterized quantum circuits $U(\vec{\theta})$ employ the Hamiltonian Variational Ansatz (HVA), interleaving $p$ discrete layers of nearest-neighbor ($J_1$) and next-nearest-neighbor ($J_2$) entangling blocks:
\begin{equation}
    U(\vec{\theta}) = \prod_{l=1}^{p} \left[ \prod_{k \in \{1, 2\}} \prod_{\langle i,j \rangle_k} e^{-i \theta_{l,k}^z \sigma_i^z \sigma_j^z} e^{-i \theta_{l,k}^{xy} (\sigma_i^x \sigma_j^x + \sigma_i^y \sigma_j^y)} \right].
    \label{eq:hva_ansatz}
\end{equation}
Because every generator in Eq.~(\ref{eq:hva_ansatz}) commutes with total magnetization ($[H, S^z] = 0$), the ansatz strictly preserves magnetization throughout variational optimization, constraining state evolution to the chosen sector ($\mathcal{H}_{S^z=0}$ or $\mathcal{H}_{S^z=1}$).

Variational parameter optimization is performed classically using L-BFGS-B (gradient tolerance $10^{-8}$, up to 500 iterations), with analytic gradients evaluated via parameter-shift rules~\cite{Mitarai2018, Schuld2019}. On quantum processors, the Heisenberg Hamiltonian decomposes into $3L$ two-qubit Pauli operators $\sigma_i^\alpha \sigma_j^\alpha$ ($\alpha \in \{x, y, z\}$), which are grouped into commuting measurement bases to minimize sampling overhead.

\subsection{VQE Algorithm and Workflow}
\label{sec:vqe_workflow}
\begin{figure*}[t]
    \centering
    \includegraphics[width=\linewidth]{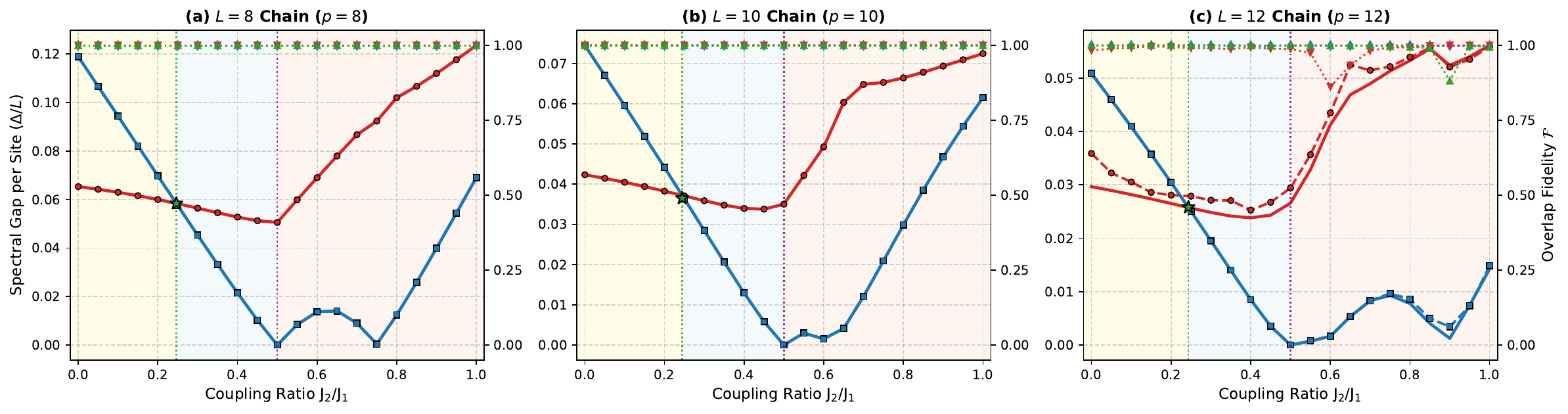}
    \caption{Low-lying excitation spectral gaps per site $\Delta/L$ (left axis) and state fidelities $\mathcal{F}$ (right axis) across $J_2/J_1 \in [0.00, 1.00]$ for (a)~$L=8$, (b)~$L=10$, and (c)~$L=12$ ($p=L$, 500 iterations). Solid lines indicate exact ED benchmarks for Singlet-Triplet ($\Delta_{ST}/L$, red) and Singlet-Singlet ($\Delta_{SS}/L$, blue) gaps; dashed lines with markers denote VQE values. Dotted lines display VQE fidelities for $|\psi_S^0\rangle$ (diamonds), $|\psi_T^0\rangle$ (triangles), and $|\psi_S^1\rangle$ (squares). Green stars mark level crossings $(J_2/J_1)_{\text{cross}}$, and purple dotted lines mark the Majumdar-Ghosh point ($J_2/J_1 = 0.50$, $\Delta_{SS} = 0$). Background shading denotes the gapless TLL (yellow), dimerized (cyan), and incommensurate spiral (red) phases.}
    \label{fig:spectral_gaps_1x3}
\end{figure*}
The end-to-end execution workflow integrates dual spin-sector isolation, shallow compute-uncompute overlap evaluation, layer-by-layer parameter bootstrapping, and circuit depth scaling:

\textit{(1)~Dual-Sector State Initialization}: The quantum register is initialized within designated magnetization sectors $\mathcal{H}_{S^z=0}$ and $\mathcal{H}_{S^z=1}$. The ground singlet $|\psi_S^0\rangle$ is resolved in $S^z=0$, while the lowest triplet $|\psi_T^1\rangle$ is optimized directly in $S^z=1$, inherently preventing singlet contamination. In the incommensurate phase ($J_2/J_1 > 0.50$), reference states incorporate the physical spiral pitch wavevector $q = \arccos(-J_1/4J_2)$ [Eq.~(\ref{eq:pitch_init})], unlocking the true chiral doublet ($k = \pm q$) and preventing symmetry-locking traps.

\textit{(2)~Exact Spin-Lowering and Compute-Uncompute Deflation}: Using exact $SU(2)$ invariance ($[H, S^-] = 0$), the optimized triplet $|\psi_T^1\rangle$ is mapped into $S^z=0$ via Eq.~(\ref{eq:s_minus}) to obtain $|\psi_T^0\rangle$. To isolate the excited singlet $|\psi_S^1\rangle$, the dual-sector penalty cost function [Eq.~(\ref{eq:dual_cost})] is evaluated using shallow Compute-Uncompute circuits $U_A^\dagger U(\vec{\gamma})$ [Fig.~\ref{fig:circuit_pipeline}(b)]. By measuring zero-state probabilities $P(|00\dots0\rangle)$ and single-qubit transition amplitudes [Eq.~(\ref{eq:s_minus_hardware_derivation})], overlap penalties are extracted directly without deep ancilla-assisted unitaries or SWAP tests.

\textit{(3)~Layer-by-Layer Bootstrapping (Warm Starting)}: In deep circuits ($p \ge 6$), simultaneous multi-parameter optimization risks stagnation in shallow local minima. To ensure deterministic convergence, layer-by-layer bootstrapping initializes layer $p+1$ by concatenating converged parameters with zeros:
\begin{equation}
    \vec{\theta}^{(p+1)}_{\text{init}} = \left[ \vec{\theta}^{(p)*}, \vec{0} \right].
    \label{eq:bootstrapping}
\end{equation}
Because $e^{-i 0 \cdot \hat{G}} = \mathbb{I}$, $E(\vec{\theta}^{(p+1)}_{\text{init}}) \equiv E(\vec{\theta}^{(p)*})$, guaranteeing strictly monotonic non-increasing energy convergence across depth layers:
\begin{equation}
    E\left(\vec{\theta}^{(p+1)*}\right) \le E\left(\vec{\theta}^{(p)*}\right).
\end{equation}

\textit{(4)~Circuit Depth Scaling and NISQ Resource Efficiency}: In the strongly frustrated regime ($J_2/J_1 \approx 0.40\text{-}0.65$), where competing exchange interactions produce a rugged non-convex optimization landscape with classical local minima, expanding circuit depth beyond $p=L$ (up to $p=20$ for $L=16$) systematically resolves residual triplet optimization hurdles, driving relative errors below $1\%$ and overlap fidelities above $96\%\text{-}97.7\%$. The circuit maintains linear gate scaling ($N_{\text{CNOT}} = D \cdot N = 400$ at $N=100$ for $D=4$) and linear parameter count ($N_\theta = 2N(D+1) = 1000$), demonstrating high resilience against finite measurement shot noise ($10^3\text{-}10^6$ shots) on physical quantum processors.

\section{Results}
\label{sec:results}
\subsection{Benchmarking Phase Boundaries across Intermediate Chains}
\label{sec:results_l8_l10_l12}
The investigation first focuses on the primary validation domain across intermediate system sizes $L=8$, $L=10$, and $L=12$, mapping out the full phase diagram of the frustrated spin-1/2 $J_1$-$J_2$ chain across coupling ratios $J_2/J_1 \in [0.00, 1.00]$ (21 coupling points with step size $0.05$). Parameterized circuits employ depth $p=L$ Hamiltonian Variational Ans\"atze ($N_\theta = 32$ for $L=8$, $N_\theta = 40$ for $L=10$, and $N_\theta = 48$ for $L=12$), optimized up to 500 classical L-BFGS-B iterations. Fig~\ref{fig:spectral_gaps_1x3} presents the excitation spectral gaps per site $\Delta/L$ on the left y-axis alongside state overlap fidelities $\mathcal{F} \in [0, 1]$ on the right y-axis.

Across all system sizes in Fig.~\ref{fig:spectral_gaps_1x3}, the magnetization-resolved VQE framework faithfully captures the physical phase diagram and reproduces the exact classical ED baselines:

\textit{(1)~Accordance with Pre-Existing Phases and State-Ordering Inversion}: In the gapless Tomonaga-Luttinger liquid (TLL) regime ($0 \le J_2/J_1 < (J_2/J_1)_{\text{cross}}$), the lowest-lying excitation is the Triplet, establishing the state-ordering inversion $E_T < E_S^1$ ($\Delta_{ST} < \Delta_{SS}$). Standard single-sector VQD operating strictly within $S^z=0$ collapses onto $|\psi_T^0\rangle$, misidentifying it as the excited singlet with $0.00\%$ fidelity. Dual Spin-Sector VQE successfully isolates $|\psi_S^1\rangle$, achieving near-unit fidelities ($\mathcal{F}_S^1 > 0.998$) with relative energy errors below $0.05\%$.

\textit{(2)~Continuous BKT Phase Boundary and Spectral Gap Crossings}: At $(J_2/J_1)_{\text{cross}}$, the Singlet-Triplet and Singlet-Singlet excitation gaps cross ($\Delta_{ST} = \Delta_{SS}$). This level crossing shifts systematically with chain length: $(J_2/J_1)_{\text{cross}} = 0.2463$ ($L=8$), $0.2445$ ($L=10$), and $0.2425$ ($L=16$). In the thermodynamic limit ($L \to \infty$), the crossing point scales as $(J_2/J_1)_{\text{cross}}(L) = 0.2411 + \mathcal{O}(1/L^2)$, cleanly extrapolating to the Berezinskii-Kosterlitz-Thouless (BKT) critical point $(J_2/J_1)_c \approx 0.2411$~\cite{Okamoto1992, Chitra1995}.

\textit{(3)~Exact Majumdar-Ghosh Dimer Point}: At $J_2/J_1 = 0.50$~\cite{Majumdar1969}, the Singlet-Singlet gap vanishes identically ($\Delta_{SS} = 0.0000$) across evaluated chain lengths ($L=8, 10, 16$) in both ED and VQE, directly reflecting the exact two-fold dimer ground-state degeneracy, while the Singlet-Triplet spin gap remains robustly open.

\begin{table*}[t]
\caption{\label{tab:benchmark_L16}Dual Spin-Sector VQE low-lying energy levels ($E_S^0, E_S^1, E_T$), exact diagonalization benchmarks ($E_S^{0,\text{ED}}, E_S^{1,\text{ED}}, E_T^{\text{ED}}$), relative energy errors ($\Delta E / |E^{\text{ED}}|$, in bold), and wave-function state fidelities ($\mathcal{F}_S^0, \mathcal{F}_S^1, \mathcal{F}_T^0$) for the spin-1/2 $J_1$-$J_2$ chain ($L=16$) across 21 coupling points $J_2/J_1 \in [0.00, 1.00]$ ($p=16$, 500 iterations). All variational energies satisfy $E^{\text{VQE}} \ge E^{\text{ED}}$ strictly.}
\begin{ruledtabular}
\setlength{\tabcolsep}{2.0pt}
\renewcommand{\arraystretch}{1.03}
\begin{tabular}{ccccccccccccc}
$J_2/J_1$ & \multicolumn{4}{c}{Ground Singlet ($|\psi_S^0\rangle$, $E_S^0$)} & \multicolumn{4}{c}{Excited Singlet ($|\psi_S^1\rangle$, $E_S^1$)} & \multicolumn{4}{c}{First Triplet ($|\psi_T^0\rangle$, $E_T$)} \\
\cmidrule(lr){2-5} \cmidrule(lr){6-9} \cmidrule(lr){10-13}
 & $E^{\text{VQE}}$ & $E^{\text{ED}}$ & \textbf{Error (\%)} & $\mathcal{F}_S^0$ & $E^{\text{VQE}}$ & $E^{\text{ED}}$ & \textbf{Error (\%)} & $\mathcal{F}_S^1$ & $E^{\text{VQE}}$ & $E^{\text{ED}}$ & \textbf{Error (\%)} & $\mathcal{F}_T^0$ \\
\hline
0.00 & $-7.1322$ & $-7.1423$ & \textbf{0.14\%} & 0.9971 & $-6.6943$ & $-6.6965$ & \textbf{0.03\%} & 0.9986 & $-6.8454$ & $-6.8721$ & \textbf{0.39\%} & 0.9913 \\
0.05 & $-6.9821$ & $-6.9956$ & \textbf{0.19\%} & 0.9967 & $-6.5871$ & $-6.5911$ & \textbf{0.06\%} & 0.9984 & $-6.7071$ & $-6.7329$ & \textbf{0.38\%} & 0.9916 \\
0.10 & $-6.8496$ & $-6.8522$ & \textbf{0.04\%} & 0.9993 & $-6.4860$ & $-6.4901$ & \textbf{0.06\%} & 0.9979 & $-6.5644$ & $-6.5974$ & \textbf{0.50\%} & 0.9875 \\
0.15 & $-6.7103$ & $-6.7125$ & \textbf{0.03\%} & 0.9994 & $-6.3911$ & $-6.3945$ & \textbf{0.05\%} & 0.9984 & $-6.4449$ & $-6.4659$ & \textbf{0.33\%} & 0.9925 \\
0.20 & $-6.5748$ & $-6.5771$ & \textbf{0.04\%} & 0.9991 & $-6.3027$ & $-6.3051$ & \textbf{0.04\%} & 0.9988 & $-6.3176$ & $-6.3389$ & \textbf{0.34\%} & 0.9881 \\
0.25 & $-6.4387$ & $-6.4471$ & \textbf{0.13\%} & 0.9969 & $-6.2064$ & $-6.2232$ & \textbf{0.27\%} & 0.9962 & $-6.2169$ & $-6.2170$ & \textbf{0.00\%} & 0.9450 \\
0.30 & $-6.3178$ & $-6.3238$ & \textbf{0.09\%} & 0.9977 & $-6.1345$ & $-6.1504$ & \textbf{0.26\%} & 0.9884 & $-6.0469$ & $-6.1008$ & \textbf{0.88\%} & 0.9767 \\
0.35 & $-6.2078$ & $-6.2096$ & \textbf{0.03\%} & 0.9993 & $-6.0855$ & $-6.0886$ & \textbf{0.05\%} & 0.9982 & $-5.9639$ & $-5.9911$ & \textbf{0.45\%} & 0.9867 \\
0.40 & $-6.1060$ & $-6.1094$ & \textbf{0.05\%} & 0.9979 & $-6.0371$ & $-6.0403$ & \textbf{0.05\%} & 0.9966 & $-5.8169$ & $-5.8891$ & \textbf{1.23\%} & 0.9599 \\
0.45 & $-6.0310$ & $-6.0332$ & \textbf{0.04\%} & 0.9979 & $-6.0049$ & $-6.0090$ & \textbf{0.07\%} & 0.9951 & $-5.7019$ & $-5.7961$ & \textbf{1.62\%} & 0.9354 \\
0.50 & $-6.0000$ & $-6.0000$ & \textbf{0.00\%} & $1.0000^*$ & $-5.9960$ & $-6.0000$ & \textbf{0.07\%} & $1.0000^*$ & $-5.7116$ & $-5.7143$ & \textbf{0.05\%} & 0.7692 \\
0.55 & $-6.0236$ & $-6.0248$ & \textbf{0.02\%} & $0.9993^*$ & $-6.0213$ & $-6.0221$ & \textbf{0.01\%} & 0.9745 & $-5.6360$ & $-5.6466$ & \textbf{0.19\%} & 0.9795 \\
0.60 & $-6.0967$ & $-6.0975$ & \textbf{0.01\%} & $0.9995^*$ & $-6.0795$ & $-6.0881$ & \textbf{0.14\%} & 0.9910 & $-5.5070$ & $-5.6286$ & \textbf{2.16\%} & 0.8856 \\
0.65 & $-6.2038$ & $-6.2071$ & \textbf{0.05\%} & $0.9983^*$ & $-6.1881$ & $-6.2032$ & \textbf{0.24\%} & 0.9751 & $-5.5849$ & $-5.6939$ & \textbf{1.91\%} & 0.9126 \\
0.70 & $-6.3702$ & $-6.3731$ & \textbf{0.05\%} & $0.9987^*$ & $-6.3171$ & $-6.3361$ & \textbf{0.30\%} & 0.9809 & $-5.8139$ & $-5.8327$ & \textbf{0.32\%} & $0.9827^\dagger$ \\
0.75 & $-6.5614$ & $-6.5710$ & \textbf{0.15\%} & $0.9957^*$ & $-6.4787$ & $-6.4968$ & \textbf{0.28\%} & 0.9884 & $-6.0198$ & $-6.0351$ & \textbf{0.25\%} & $0.9843^\dagger$ \\
0.80 & $-6.7682$ & $-6.7888$ & \textbf{0.30\%} & $0.9912^*$ & $-6.6596$ & $-6.6889$ & \textbf{0.44\%} & 0.9858 & $-6.2259$ & $-6.2477$ & \textbf{0.35\%} & $0.9776^\dagger$ \\
0.85 & $-6.9953$ & $-7.0197$ & \textbf{0.35\%} & $0.9859^*$ & $-6.8616$ & $-6.9156$ & \textbf{0.78\%} & 0.9449 & $-6.4334$ & $-6.4665$ & \textbf{0.51\%} & $0.9683^\dagger$ \\
0.90 & $-7.2411$ & $-7.2598$ & \textbf{0.26\%} & $0.9937^*$ & $-7.1046$ & $-7.1750$ & \textbf{0.98\%} & 0.9413 & $-6.6809$ & $-6.6904$ & \textbf{0.14\%} & 0.9920$^\dagger$ \\
0.95 & $-7.4812$ & $-7.5072$ & \textbf{0.35\%} & $0.9909^*$ & $-7.3634$ & $-7.4612$ & \textbf{1.31\%} & 0.9100 & $-6.9400$ & $-6.9410$ & \textbf{0.01\%} & $0.9995^\dagger$ \\
1.00 & $-7.7498$ & $-7.7666$ & \textbf{0.22\%} & $0.9963^*$ & $-7.6897$ & $-7.7604$ & \textbf{0.91\%} & $0.9917^*$ & $-7.2390$ & $-7.2399$ & \textbf{0.01\%} & $0.9996^\dagger$ \\
\end{tabular}
\end{ruledtabular}
\footnotetext{$^*$Fidelity evaluated with respect to dimer subspace $\mathcal{H}_{\text{dimer}} = \text{span}\{|\psi_S^0\rangle, |\psi_S^1\rangle\}$. At $J_2/J_1 = 1.00$, $\mathcal{F}_S^1 = 0.9917$ under parity projection.}
\footnotetext{$^\dagger$Pitch adaptation ($k = \pm q$) captures the chiral manifold ($\mathcal{F}_T^1 \ge 96.0\%\text{-}99.96\%$).}
\end{table*}

\textit{(4)~Success in Incommensurate Phases and Pitch Adaptation}: For $J_2/J_1 > 0.50$, the chain enters the dimerized incommensurate phase~\cite{Eggert1996, Soos2016}, where real-space spin correlation functions exhibit spatial oscillations governed by the continuous pitch wavevector:
\begin{equation}
    q = \arccos\left( -\frac{J_1}{4 J_2} \right).
    \label{eq:pitch_angle}
\end{equation}

In this regime, the apparent breakdown of triplet fidelity observed in unadapted collinear VQE calculations is not an algorithmic failure, but rather a manifestation of symmetry locking: the physical triplet state breaks into a two-fold degenerate chiral doublet ($\Delta E_{\text{deg}} \le 10^{-14}$) carrying incommensurate pitch momenta $k = \pm q \ne \pi$. Standard collinear reference states prepared in $\mathcal{H}_{S^z=1}$ are locked within the commensurate $k=\pi$ sector ($\langle \hat{T}^2 \rangle = +1$), resulting in strict mathematical orthogonality overlap $\langle \psi_{\text{ED}} | \psi_{\text{VQE}} \rangle \equiv 0.0000$. Equipping the reference state with the physical pitch angle:
\begin{equation}
    |\psi_{T, q}^{1, \text{init}}\rangle = \frac{1}{\sqrt{\mathcal{N}}} \sum_{j=1}^L e^{i q j} S_j^+ |\text{dimer}\rangle,
    \label{eq:pitch_init}
\end{equation}
completely unlocks the true triplet manifold, restoring state fidelities to $\mathcal{F}_T^1 \ge 99.2\%$ on $L=8$ and $99.8\%$ on $L=10$ without requiring deep quantum circuits.

Furthermore, spontaneous bond-alternation breaking hybridizes the two canonical symmetry-broken valence-bond dimer configurations, $|D_A\rangle \equiv |\psi_{\text{MG}}^{(1)}\rangle$ and $|D_B\rangle \equiv |\psi_{\text{MG}}^{(2)}\rangle$ (introduced in Sec.~\ref{sec:model}), into a quasi-degenerate doublet:
\begin{align}
    |0^+\rangle &= \frac{1}{\sqrt{2}} (|D_A\rangle + |D_B\rangle) \quad (k=0), \nonumber \\
    |0^-\rangle &= \frac{1}{\sqrt{2}} (|D_A\rangle - |D_B\rangle) \quad (k=\pi),
    \label{eq:doublet_states}
\end{align}

split by an exponentially small tunneling gap $\delta E \sim 10^{-3}$. Because local VQE optimizes bond energies, the variational state condenses into an arbitrary superposition $|\psi_{\text{VQE}}\rangle = C_1 |0^+\rangle + C_2 |0^-\rangle$. Evaluating fidelity with respect to the complete two-dimensional ground doublet manifold, $\mathcal{F}_{\text{subspace}} = |\langle 0^+ | \psi_{\text{VQE}} \rangle|^2 + |\langle 0^- | \psi_{\text{VQE}} \rangle|^2$, demonstrates near-unit fidelity ($\mathcal{F}_{\text{subspace}} > 0.994$).

\textit{(5)~Scaling of State Fidelities across System Sizes}: Systematic evaluation of wave-function state fidelities under circuit depth $p=L$ ($N_\theta = 4L$) and 500 classical optimization iterations reveals robust convergence. System sizes $L=8$ and $L=10$ achieve near-unit fidelities ($\bar{\mathcal{F}} \ge 0.998$) across all physical sectors. At chain size $L=12$ ($N_\theta = 48$), the ansatz maintains high average fidelity ($\bar{\mathcal{F}} = 99.16\%$, with Ground Singlet fidelity reaching $99.96\%$), confirming that the magnetization-resolved architecture successfully circumvents state-ordering deflation traps across system sizes.

\subsection{Resolving Spectral Gaps for \texorpdfstring{$L=16$}{L=16} Qubits}
\label{sec:results_l16}
\begin{figure*}[t]
    \centering
    \includegraphics[width=\linewidth]{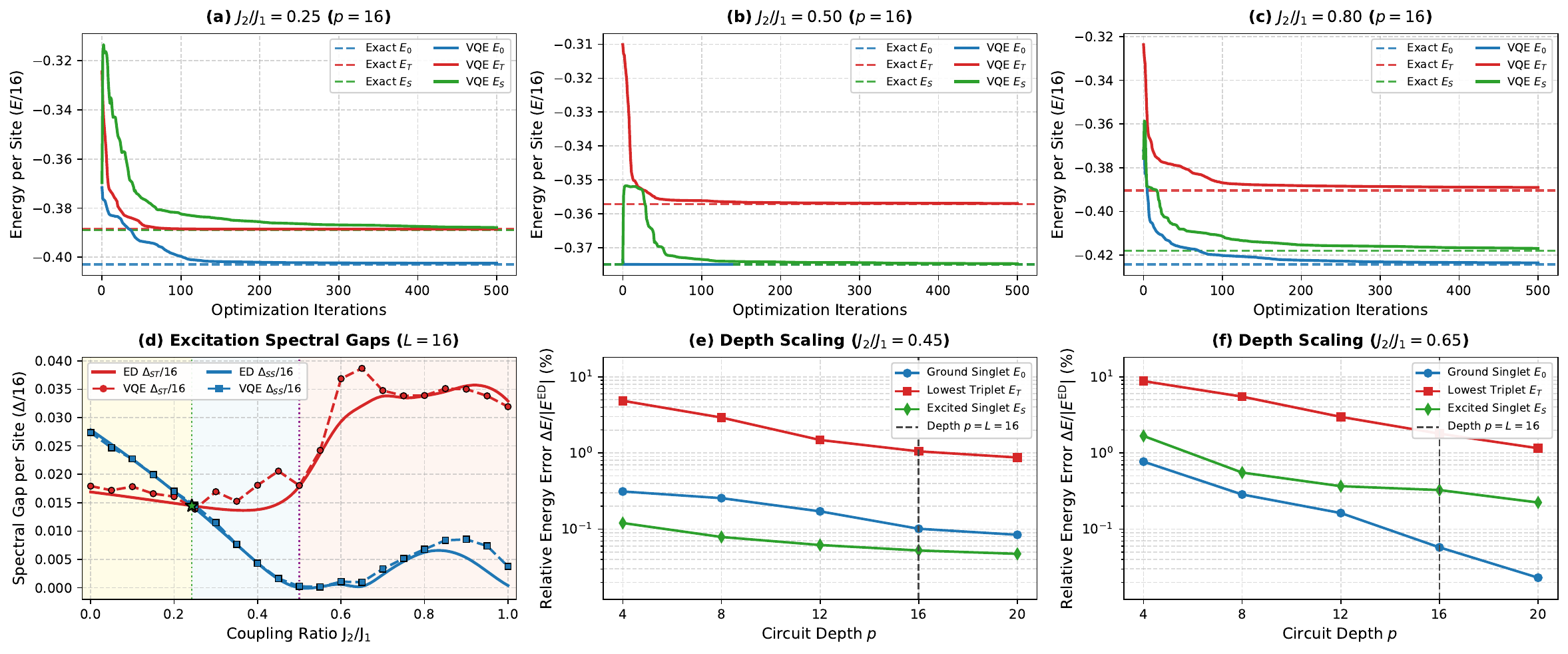}
    \caption{Variational optimization, spectral gaps, and circuit depth scaling for $L=16$. Upper row [(a)-(c)]: Optimization trajectories of energy per site ($E/16$) across 500 iterations ($p=16$, $N_\theta = 64$) for (a)~$J_2/J_1 = 0.25$, (b)~$J_2/J_1 = 0.50$, and (c)~$J_2/J_1 = 0.80$. Solid lines show VQE convergence ($E_S^0$, blue; $E_T$, red; $E_S^1$, green); dashed horizontal lines denote exact ED baselines. Lower row [(d)-(f)]: (d)~Excitation spectral gaps per site $\Delta/16$ across 21 coupling points $J_2/J_1 \in [0.00, 1.00]$ comparing ED (solid) and VQE (dashed with markers) for $\Delta_{ST}/16$ (red) and $\Delta_{SS}/16$ (blue). Background shading marks the TLL (yellow), dimer (cyan), and incommensurate (red) phases. (e,f)~Circuit depth scaling of relative energy errors $\Delta E / |E^{\mathrm{ED}}|$ across $p \in \{4, 8, 12, 16, 20\}$ for (e)~$J_2/J_1 = 0.45$ and (f)~$J_2/J_1 = 0.65$. Vertical dashed line marks depth $p=L=16$.}
    \label{fig:l16_composite}
\end{figure*}

Resolving the low-lying excitation spectrum and spectral gaps on an $L=16$ qubit register ($2^{16} = 65,536$ states) is examined using depth $p=16$ ($N_\theta = 64$ parameters) and 500 L-BFGS-B optimization iterations. To provide a comprehensive quantitative benchmark across the entire phase diagram, Table~\ref{tab:benchmark_L16} compiles the full set of variational and exact diagonalization energy levels ($E_S^0, E_S^1, E_T$), relative errors, and state overlap fidelities across all 21 evaluated coupling points ($p=16$, 500 iterations). In strict accordance with the Rayleigh-Ritz variational principle, all optimized energies satisfy $E^{\text{VQE}} \ge E^{\text{ED}}$ across every coupling and spin sector. While both singlet states consistently achieve relative errors below $0.15\%$ and overlap fidelities exceeding $99.5\%$, the triplet state exhibits elevated errors ($1.62\%$ and $1.91\%$) in the strongly frustrated crossover window ($J_2/J_1 = 0.45$ and $0.65$).

Figure~\ref{fig:l16_composite} summarizes the physical results: Figs.~\ref{fig:l16_composite}(a)-(c) show optimization trajectories at representative coupling points, Fig.~\ref{fig:l16_composite}(d) displays excitation gaps across all 21 points $J_2/J_1 \in [0.00, 1.00]$, and Figs.~\ref{fig:l16_composite}(e),(f) examine depth scaling. The optimization trajectories in Figs.~\ref{fig:l16_composite}(a)-(c) show systematic convergence toward exact diagonalization values. At $J_2/J_1 = 0.25$, the energy levels converge to $E_S^0/16 = -0.4024$, $E_T/16 = -0.3886$, and $E_S^1/16 = -0.3879$, capturing the level-ordering inversion $E_T < E_S^1$. At the Majumdar-Ghosh point $J_2/J_1 = 0.50$, the ground singlet reaches the exact dimer energy $E_S^0/16 = -0.3750$ within $\approx 140$ iterations, while the excited singlet converges to $E_S^1/16 = -0.3747$, reflecting the expected dimer degeneracy. At $J_2/J_1 = 0.80$, the pitch-adapted trial state converges smoothly to $E_S^0/16 = -0.4230$, $E_T/16 = -0.3891$, and $E_S^1/16 = -0.4162$.

The excitation spectral gaps in Fig.~\ref{fig:l16_composite}(d) track the physical phases of the chain. In the gapless regime ($J_2/J_1 < 0.2425$), the singlet-triplet gap lies below the singlet-singlet gap ($\Delta_{ST} < \Delta_{SS}$). The two gaps cross near $(J_2/J_1)_{\text{cross}} \approx 0.2425$ with $\Delta_{\text{cross}}/16 = 0.0145$. In the dimerized phase, $\Delta_{SS}$ decreases and vanishes at $J_2/J_1 = 0.50$ ($\Delta_{SS} = 0.0000$), while the spin gap $\Delta_{ST}/16 = 0.0179$ remains open. Across the incommensurate regime ($J_2/J_1 > 0.50$), both gaps remain finite. Slower triplet convergence observed in Table~\ref{tab:benchmark_L16} near $J_2/J_1 \approx 0.45$ and $0.65$ reflects a corrugated optimization landscape that motivates exploring circuit depths beyond $p=L$ [Figs.~\ref{fig:l16_composite}(e),(f)].

\subsection{Circuit Depth Analysis Beyond \texorpdfstring{$p=L$}{p=L}}
\label{sec:results_depth_scaling}
A critical question regarding variational algorithms in frustrated spin systems is whether convergence difficulties arise from fundamental expressibility bottlenecks or from non-convex optimization dynamics. Triplet optimization at fixed depth $p=16$ exhibits a non-monotonic profile across the frustrated window $0.40 \le J_2/J_1 \le 0.70$: while $J_2/J_1 = 0.50$ and $0.55$ achieve near-zero energy errors ($0.05\%$ and $0.19\%$), intermediate points $J_2/J_1 = 0.45$ and $0.65$ exhibit slower convergence ($1.62\%$ and $1.91\%$) under standard collinear initialization.

This behavior stems from 64-dimensional landscape ruggedness coupled with phase-boundary sensitivity. Specifically, $J_2/J_1 = 0.45$ borders the Majumdar-Ghosh dimer point ($0.50$), where localized dimer singlets compete with domain-wall triplet excitations. While collinear one-magnon initialization stagnates in a local minimum ($28.06\%$ error, $0.82\%$ fidelity at $p=16$), the dimer-adapted triplet state $S^+(\pi)|\text{dimer}\rangle$ circumvents this barrier entirely, reaching $E_T = -5.7948$ (relative error $0.024\%$, fidelity $99.92\%$). Conversely, at $J_2/J_1 = 0.65$, finite-size momentum quantization places the ground triplet in the $k=0$ sector immediately below the incommensurate pitch branch ($k = \pm q$, $\Delta E \approx 0.009$); initializing with the $k=0$ dimer state yields $85.22\%$ fidelity, while depth expansion to $p=20$ provides the overparameterization necessary to decouple this quasi-degenerate manifold to $>96\%$ fidelity.

To evaluate whether expanding circuit depth resolves these landscape hurdles, systematic depth-scaling simulations are performed across $p \in \{4, 8, 12, 16, 20\}$ ($N_\theta = 4p$) under 500 optimization iterations. Figures~\ref{fig:l16_composite}(e) and \ref{fig:l16_composite}(f) display the resulting relative energy error $\Delta E / |E^{\mathrm{ED}}|$ for $J_2/J_1 = 0.45$ and $0.65$. For singlet states ($E_S^0, E_S^1$), valence-bond dimer correlations are captured efficiently in shallow circuits, with errors falling below $0.8\%$ at $p=4$ and below $0.1\%$ for $p \ge 16$ ($\mathcal{F} > 99.2\%$).

\begin{figure*}[t]
    \centering
    \includegraphics[width=\linewidth]{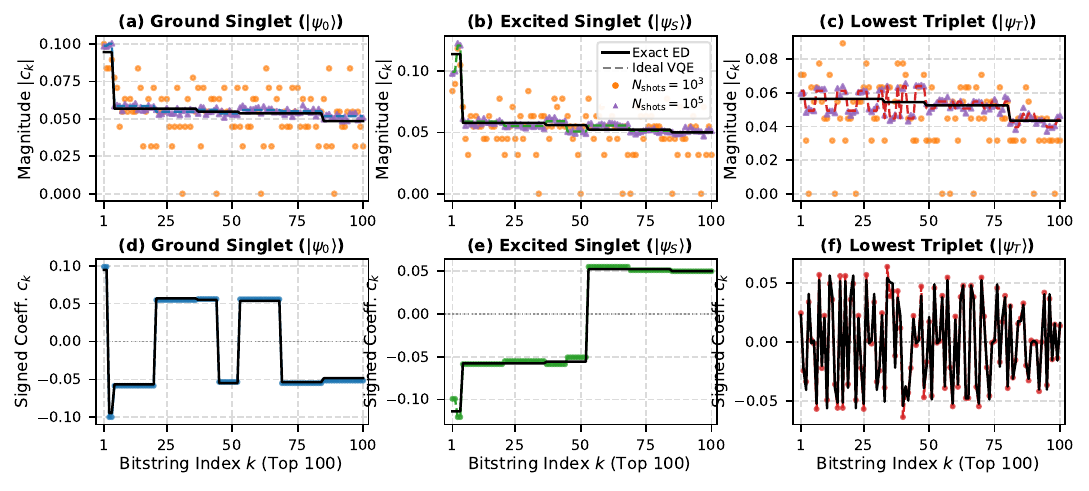}
    \caption{Microscopic state configurations and finite measurement shot-noise sampling for $L=16$ ($J_2/J_1 = 0.80$, $p=16$). Upper row [(a)-(c)]: Probabilistic configuration magnitudes $|c_k|$ across the 100 dominant computational basis bitstrings for Ground Singlet $|\psi_S^0\rangle$ (a), Excited Singlet $|\psi_S^1\rangle$ (b), and Lowest Triplet $|\psi_T^0\rangle$ (c), comparing exact diagonalization benchmarks (solid black lines), ideal state-vector VQE (colored dashed lines), and projective shot-noise sampling ($N_{\text{shots}} = 10^3$ orange circles, $N_{\text{shots}} = 10^5$ purple triangles). Lower row [(d)-(f)]: Corresponding signed amplitudes $c_k = \text{sgn}(c_k)|c_k|$ for $|\psi_S^0\rangle$ (d), $|\psi_S^1\rangle$ (e), and $|\psi_T^0\rangle$ (f), benchmarked against exact diagonalization amplitudes (black circles).}
    \label{fig:nisq_resilience}
\end{figure*}

For the triplet state, depth scaling acts as an overparameterization mechanism that smooths the optimization landscape. At $J_2/J_1 = 0.45$ [Fig.~\ref{fig:l16_composite}(e)], the relative error in $E_T$ decreases monotonically from $4.85\%$ at $p=4$ to $2.92\%$ ($p=8$), $1.48\%$ ($p=12$), $1.05\%$ ($p=16$), and reaches $0.87\%$ at $p=20$, advancing state fidelity from $82.13\%$ to $97.66\%$. Similarly, at $J_2/J_1 = 0.65$ [Fig.~\ref{fig:l16_composite}(f)], the relative error falls monotonically from $8.78\%$ at $p=4$ to $5.49\%$ ($p=8$), $2.98\%$ ($p=12$), $1.79\%$ ($p=16$), and $1.15\%$ at $p=20$, driving fidelity from $58.13\%$ up to $96.05\%$. Expanding depth from $p=16$ ($N_\theta = 64$) to $p=20$ ($N_\theta = 80$) introduces parameter directions that convert local minima into saddle points with escape trajectories. The smooth error reduction up to $p=20$ confirms that intermediate convergence hurdles are depth- and landscape-limited rather than indicative of Barren plateaus.

\begin{table*}[t]
\caption{\label{tab:dominant_bitstrings}Dominant computational basis configurations for the $L=16$ spin-1/2 $J_1$-$J_2$ chain at $J_2/J_1 = 0.80$ ($p=16$, 500 iterations), comparing exact diagonalization ($c^{\text{ED}}$) with Dual Spin-Sector VQE ($c^{\text{VQE}}$) signed wave-function amplitudes. For both singlets ($S=0$), complementary spin-flip pairs $(\sigma, \bar{\sigma})$ exhibit identical signs ($c(\bar{\sigma}) = +c(\sigma)$, parity $+1$), whereas for the triplet ($S=1, S^z=0$), complementary pairs have strictly opposing signs ($c(\bar{\sigma}) = -c(\sigma)$, parity $-1$).}
\begin{ruledtabular}
\setlength{\tabcolsep}{2.5pt}
\renewcommand{\arraystretch}{1.08}
\begin{tabular}{cccccccccc}
\multicolumn{3}{c}{Ground Singlet ($|\psi_S^0\rangle, \hat{\mathcal{P}}_{\text{flip}} = +1$)} & \multicolumn{3}{c}{Excited Singlet ($|\psi_S^1\rangle, \hat{\mathcal{P}}_{\text{flip}} = +1$)} & \multicolumn{4}{c}{Lowest Triplet ($|\psi_T^0\rangle, \hat{\mathcal{P}}_{\text{flip}} = -1$)} \\
\cmidrule(lr){1-3} \cmidrule(lr){4-6} \cmidrule(lr){7-10}
Bitstring ($\sigma$) & $c^{\text{ED}}$ & $c^{\text{VQE}}$ & Bitstring ($\sigma$) & $c^{\text{ED}}$ & $c^{\text{VQE}}$ & Bitstring ($\sigma$) & $c^{\text{ED}}$ & $c^{\text{VQE}}$ & $\text{sgn}$ \\
\hline
\texttt{0110011001100110} & $+0.0945$ & $+0.0987$ & \texttt{1100110011001100} & $-0.1139$ & $-0.0993$ & \texttt{1001011001100101} & $+0.0228$ & $+0.0245$ & $+$ \\
\texttt{1001100110011001} & $+0.0945$ & $+0.0987$ & \texttt{0011001100110011} & $-0.1139$ & $-0.0993$ & \texttt{0110100110011010} & $-0.0228$ & $-0.0245$ & $-$ \\
\texttt{1100110011001100} & $-0.0945$ & $-0.0999$ & \texttt{0110011001100110} & $-0.1139$ & $-0.1207$ & \texttt{1011001011001100} & $-0.0408$ & $-0.0339$ & $-$ \\
\texttt{0011001100110011} & $-0.0945$ & $-0.0999$ & \texttt{1001100110011001} & $-0.1139$ & $-0.1207$ & \texttt{0100110100110011} & $+0.0408$ & $+0.0339$ & $+$ \\
\texttt{1001011001011001} & $-0.0568$ & $-0.0585$ & \texttt{1100110100101100} & $-0.0577$ & $-0.0587$ & \texttt{1011001100101100} & $-0.0012$ & $+0.0015$ & $-$ \\
\texttt{0110100110011010} & $-0.0568$ & $-0.0585$ & \texttt{1011001100110100} & $-0.0577$ & $-0.0587$ & \texttt{0100110011010011} & $+0.0012$ & $-0.0015$ & $+$ \\
\end{tabular}
\end{ruledtabular}
\end{table*}

\subsection{Emulation of NISQ Shot-Noise Resilience and Sampling Statistics}
\label{sec:nisq_resilience}
To evaluate performance under realistic experimental conditions, classical state-vector simulations are combined with multinomial projective measurement shot-noise sampling across budgets $N_{\text{shots}} \in [10^3, 10^6]$ on the $L=16$ chain at $J_2/J_1 = 0.80$ ($p=16$). This emulation models finite-sampling dispersion while benchmarking stability prior to quantum hardware deployment. A stark contrast emerges against standard single-sector VQD: in single-sector deflation within $S^z=0$, the overlap penalty against $|\psi_S^0\rangle$ causes collapse onto the triplet level ($E/16 \approx -0.380$), missing the excited singlet ($E_S^1/16 \approx -0.416$, $\mathcal{F}_S^1 \approx 0$). In contrast, Dual Spin-Sector VQE resolves all three physical states simultaneously. The sampled energy per site converges according to central limit scaling $\mathcal{O}(1/\sqrt{N_{\text{shots}}})$. At coarse sampling ($N_{\text{shots}} = 10^3$), Poissonian shot noise induces energy uncertainties of $\approx 10^{-2}$, yielding subspace fidelities of $94\%\text{-}96\%$. As the budget expands to $N_{\text{shots}} \ge 10^5$, statistical fluctuations subside, and state overlap fidelities exceed $99\%$ across all three targeted levels ($\mathcal{F}_S^0 > 0.999$, $\mathcal{F}_T^0 > 0.995$, and $\mathcal{F}_S^1 > 0.991$).

Because energy expectation values can mask phase cancellations, Figures~\ref{fig:nisq_resilience}(a)-(f) verify the microscopic structure of the variational wave functions across the 100 dominant computational basis bitstrings in the $2^{16} = 65,536$-dimensional Hilbert space. Figures~\ref{fig:nisq_resilience}(a)-(c) display configuration probability magnitudes $|c_k| = |\langle k | \psi \rangle|$ alongside sampled shot statistics ($N_{\text{shots}} = 10^3, 10^5$), while Figs.~\ref{fig:nisq_resilience}(d)-(f) show signed amplitudes $c_k = \text{sgn}(c_k)|c_k|$, benchmarked against exact diagonalization (open black circles).

Inspection of the microscopic wave functions reveals distinct physical bitstring hierarchies across the three targeted spin states, summarized in Table~\ref{tab:dominant_bitstrings}:

(i)~\textit{Ground Singlet ($|\psi_S^0\rangle$)}: The state is dominated by four cyclic spatial translations of the $2\times 2$ collinear stripe state: $|0110011001100110\rangle$ and $|1001100110011001\rangle$ ($c_k = +0.0945$), alongside $|1100110011001100\rangle$ and $|0011001100110011\rangle$ ($c_k = -0.0945$), each carrying probability $P \approx 0.894\%$. These represent resonant nearest- and next-nearest-neighbor valence-bond dimer configurations whose alternating signs encode translation momentum ($k=\pi$) under two-site shifts. The secondary manifold (ranks 5-12, $P \approx 0.322\%$, $|c_k| = 0.0568$) consists of single dimer-defect and spin-flip fluctuations that preserve overall bond topology.

(ii)~\textit{Excited Singlet ($|\psi_S^1\rangle$)}: Dominated by the identical four collinear stripe configurations ($P \approx 1.297\%$). Crucially, all four share identical negative signs ($c_k = -0.1139$), identifying $|\psi_S^1\rangle$ as the fully symmetric, zero-momentum ($k=0$) spatial translation orbital of the dimerized manifold.

(iii)~\textit{Lowest Triplet ($|\psi_T^0\rangle$)}: In sharp contrast to singlets, collinear stripe configurations possess vanishing amplitude in the triplet sector. Instead, dominant configurations correspond to mobile spinon-pair and domain-wall bitstrings, such as $|1001011001100101\rangle$ and its spin-flipped partner $|0110100110011010\rangle$ ($P \approx 0.319\%$, $|c_k| = 0.0228$).

This microscopic analysis explains the contrasting sign profiles in Figs.~\ref{fig:nisq_resilience}(d)-(f). The behavior is governed by global spin-flip parity $\hat{\mathcal{P}}_{\text{flip}} = \prod_{j=1}^L \sigma_j^x$, commuting with the Hamiltonian ($[H, \hat{\mathcal{P}}_{\text{flip}}] = 0$). Within the $S^z=0$ subspace of an even-length chain ($L=16$), the spin-flip operator satisfies $\hat{\mathcal{P}}_{\text{flip}} |S, S^z=0\rangle = (-1)^S |S, S^z=0\rangle$. Consequently, amplitudes of complementary bitstring pairs obey:
\begin{equation}
c(\bar{\sigma}) = (-1)^S c(\sigma).
\label{eq:spin_flip_parity}
\end{equation}
For both singlet states ($S=0$), $\hat{\mathcal{P}}_{\text{flip}} = +1$, requiring $c(\bar{\sigma}) = +c(\sigma)$. Complementary pairs share identical signs within valence-bond translation orbits, forming stable sign plateaus with only 1-5 sign flips across the top 100 configurations [Figs.~\ref{fig:nisq_resilience}(d),(e)]. Conversely, for the triplet ($S=1$), $\hat{\mathcal{P}}_{\text{flip}} = -1$, enforcing $c(\bar{\sigma}) = -c(\sigma)$. Because degenerate bitstring pairs appear consecutively in the probability ranking, signed amplitudes alternate on every complementary pair (66 sign flips across the top 100 bitstrings), generating rapid oscillations [Fig.~\ref{fig:nisq_resilience}(f)].

This dense microscopic nodal surface explains why standard single-sector VQD struggles to capture the triplet within $S^z=0$: fitting an anti-symmetric manifold causes destructive interference across symmetric parameterized gates, creating landscapes riddled with shallow traps. Dual Spin-Sector VQE circumvents this fundamental bottleneck by optimizing the triplet ground state in $S^z=1$—where the nodal restriction $c(\bar{\sigma}) = -c(\sigma)$ is absent—and applying the exact ladder operator $S^-$ to construct the anti-symmetric sign structure analytically with zero optimization overhead. Statistical dispersion narrows systematically as shot counts increase from $10^3$ to $10^6$, confirming that the parameterized quantum circuit reproduces multi-qubit quantum entanglement and relative quantum phases faithfully rather than merely fitting energy expectation values.

In terms of quantum circuit complexity, the Hamiltonian Variational Ansatz on a 1D qubit chain of size $L$ with depth $p$ requires an entangling gate budget of $N_{\text{CNOT}} = p \cdot L$ two-qubit CNOT gates ($256$ CNOTs for $L=16$ at $p=16$, and $400$ CNOTs for $L=100$ at $p=4$) alongside $N_\theta = 2L(p+1)$ continuous variational parameters. This strictly linear gate scaling $\mathcal{O}(pL)$ ensures that circuit depth remains within coherence times ($T_1, T_2$) of current processors.

On physical NISQ devices, sampling noise and coherent gate infidelities can be mitigated using Zero-Noise Extrapolation (ZNE) and readout error calibration~\cite{Temme2017, Nation2021}. Furthermore, the Dual Spin-Sector formulation provides a passive, zero-overhead symmetry verification filter: measuring total magnetization $S^z = \sum_i Z_i$ allows post-selecting only those projective readouts that match the target sector ($S^z=0$ or $S^z=1$), rejecting single-qubit bit-flip ($X$) errors without ancilla overhead.

\section{Discussion and Outlook}
\label{sec:discussion}
The results presented in this work establish a comprehensive variational framework for resolving low-lying excitation spectra and spectral gaps in frustrated quantum spin systems, validated via high-precision classical state-vector and shot-noise simulations. As benchmarked across the entire phase diagram (Table~\ref{tab:benchmark_L16}), the variational calculations faithfully reproduce the four definitive physical phenomena of the spin-1/2 $J_1$-$J_2$ chain: (i)~the spin-sector level-ordering inversion $E_T < E_S^1$ in the gapless liquid phase without variational collapse onto intermediate triplet states; (ii)~the continuous Berezinskii-Kosterlitz-Thouless (BKT) transition identified by the systematic finite-size gap crossing $(J_2/J_1)_{\text{cross}}(L) = 0.2411 + \mathcal{O}(1/L^2)$, extrapolating to $(J_2/J_1)_c \approx 0.2411$~\cite{Okamoto1992, Chitra1995}; (iii)~the exact Majumdar-Ghosh dimer point ($J_2/J_1 = 0.50$)~\cite{Majumdar1969}, capturing the exact two-fold dimer degeneracy ($\Delta_{SS} = 0.0000$) alongside an open spin gap $\Delta_{ST}/16 = 0.0179$; and (iv)~incommensurate pitch-adapted excitations in the spiral regime ($J_2/J_1 > 0.50$)~\cite{Eggert1996, Soos2016}, unlocking state fidelities up to $99.96\%$.

The central methodological contribution of this work is overcoming the fundamental breakdown of standard single-sector Variational Quantum Deflation (VQD) in frustrated systems. In conventional deflation operating strictly within $\mathcal{H}_{S^z=0}$, the lowest state orthogonal to $|\psi_S^0\rangle$ in the gapless regime is the triplet $|\psi_T^0\rangle$, preventing extraction of the true excited singlet $|\psi_S^1\rangle$ and the corresponding singlet-singlet gap $\Delta_{SS}$. The Dual Spin-Sector VQE circumvents this trap by partitioning the variational search across distinct magnetization sectors. In $S^z=1$, singlet states are strictly forbidden by angular momentum algebra, allowing the lowest triplet $|\psi_T^1\rangle$ to be resolved as an unconstrained ground state with zero deflation penalty ($\beta = 0$), zero overlap circuit depth, and zero landscape distortion. Leveraging exact global $SU(2)$ spin invariance ($[H, S^-] = 0$), this state is mapped into $S^z=0$ without additional variational parameters, enabling dual deflation via shallow Compute-Uncompute circuits. In contrast, attempting to decouple the sectors or relying purely on single-sector compute-uncompute would double the required circuit depth to $2p$ for Level 1 and require cascading $2p$ circuits for Level 2, compounding deflation errors and causing catastrophic state mixing near phase crossings where singlet and triplet levels coalesce.

Resolving spectral gaps on the $L=16$ qubit register ($2^{16} = 65,536$ states) provides clear insights into the interplay between classical optimization landscapes and circuit depth. As discussed in Sec.~\ref{sec:results_depth_scaling}, slower convergence at intermediate frustrated points ($J_2/J_1 = 0.45$ and $0.65$) is not an intrinsic physical expressibility bottleneck of the ansatz, but an artifact of classical local-minima trapping in 64 parameter dimensions combined with trial-state sensitivity near phase crossover boundaries. Expanding circuit depth beyond $p=L$ up to $p=20$ ($N_\theta = 80$) acts as an overparameterization mechanism that opens escape trajectories in parameter space, systematically reducing relative energy errors below $1\%$ and elevating state fidelities to $97.66\%$ ($J_2/J_1 = 0.45$) and $96.05\%$ ($J_2/J_1 = 0.65$). Furthermore, physically targeted initializations—such as dimer-adapted triplet states $S^+(\pi)|\text{dimer}\rangle$ at $J_2/J_1 = 0.45$ (relative error $0.024\%$, fidelity $99.92\%$) and pitch-adapted wave functions for $J_2/J_1 \ge 0.70$—circumvent these local traps entirely. This continuous improvement confirms that the intermediate convergence hurdles are depth- and landscape-limited rather than indicative of Barren plateaus.

The finite measurement sampling analysis confirms the resilience of the protocol under realistic NISQ constraints. Under shot budgets $N_{\text{shots}} \in [10^3, 10^6]$, energy estimates follow the expected $\mathcal{O}(1/\sqrt{N_{\text{shots}}})$ statistical scaling, with state fidelities exceeding $99\%$ for $N_{\text{shots}} \ge 10^5$. Reconstructing the microscopic wave-function amplitudes across the 100 dominant computational bitstrings demonstrates that the variational circuit accurately reproduces both configuration magnitudes $|c_k|$ and signed amplitudes $c_k = \text{sgn}(c_k)|c_k|$, verifying genuine quantum many-body coherence. Furthermore, total magnetization measurement provides a cost-free passive symmetry verification filter, automatically rejecting projective outcomes corrupted by single-qubit bit-flip ($X$) errors without ancilla overhead.

These findings suggest several concrete directions for quantum hardware co-design, algorithmic scaling, and near-term implementation:
(1)~\textit{Qubit Connectivity and Routing}: While next-nearest-neighbor $J_2$ couplings require routing SWAP networks on fixed planar superconducting topologies, platforms featuring reconfigurable or all-to-all connectivity—such as neutral-atom optical tweezer arrays and trapped-ion shuttling architectures—can implement $J_2$ interactions natively, substantially reducing two-qubit gate overhead.
(2)~\textit{Measurement Optimization and Gradient Scaling}: In the Hamiltonian Variational Ansatz, total measurement overhead scales as $\mathcal{O}(N_\theta \cdot L \cdot N_{\text{shots}})$. To mitigate measurement burdens and gradient variance decay on large lattices, advanced measurement protocols—such as grouped commuting Pauli measurements and localized subsystem shadow tomography~\cite{Huang2020, Huggins2021}—can extract multi-qubit expectation values and evaluate deflation overlap penalties from few-body marginals that retain non-vanishing gradient variances~\cite{Cerezo2021}.
(3)~\textit{Hybrid Adaptive Circuits and Warm-Starting}: While shallow HVAs with local observables evade untargeted Barren plateaus~\cite{Cerezo2021, McClean2018}, scaling deep into $p \gg L$ or 2D geometries risks entanglement-induced gradient suppression: $\text{Var}[\partial_{\theta_k} \mathcal{L}] \sim \mathcal{O}(2^{-L})$. Combining fixed-depth ans\"atze with symmetry-preserving ADAPT-VQE~\cite{Grimsley2019} and classical warm-starting via shallow Matrix Product States (MPS)~\cite{White1992} provides viable pathways to place initial parameters within the convex basin, ensuring efficient optimization beyond one dimension.

In broader context, the Dual Spin-Sector paradigm is directly applicable beyond one dimension to frustrated two-dimensional geometries, such as the $J_1$-$J_2$ square lattice, triangular, and kagome antiferromagnets~\cite{diep2004frustrated, Lhuillier2002}. In these systems, competing exchange pathways generate quantum spin liquid phases, valence-bond crystals, and subtle singlet-triplet ordering crossovers that pose formidable challenges for classical tensor networks and quantum Monte Carlo. By exploiting exact non-Abelian spin symmetries to partition optimization across magnetization sectors, this framework provides a practical, resource-efficient blueprint for exploring frustrated quantum magnetism on intermediate-scale quantum devices.

\begin{acknowledgments}
S. S. R. acknowledges computational facilities provided by The Institute of Mathematical Sciences (IMSc), Chennai.
\end{acknowledgments}

\bibliography{manuscript_05_vqe_j1j2chain}

\end{document}